\documentclass[12pt]{article}

\usepackage{epsfig}
\usepackage{graphicx}
\usepackage{amssymb,amsmath}

\usepackage{epsfig,cite}
\usepackage{color}
\usepackage{graphics}
\usepackage{amsfonts}
\usepackage{epsfig}
\def\one{{\hbox{1\kern-.8mm l}}}

\def\non{\nonumber}

\newcommand{\beq}{\begin{equation}}
\newcommand{\eeq}{\end{equation}}
\newcommand{\be}{\begin{eqnarray}}
 \newcommand{\ee}{\end{eqnarray}}

\newcommand{\ov } {\over }

\newcommand{\s }{\sigma }

\def\a{\alpha }

\def\non{\nonumber}

\def\appendix#1{
  \addtocounter{section}{1}
  \setcounter{equation}{0}
  \renewcommand{\thesection}{\Alph{section}}
  \section*{Appendix \thesection\protect\indent \parbox[t]{11.15cm}
  {#1} }
  \addcontentsline{toc}{section}{Appendix \thesection\ \ \ #1}
  }

\begin{document}
\null\vskip-24pt 
\hfill
UB-ECM-PF-05/05
\vskip-1pt
\hfill
SISSA-19/2005/EP 
\vskip-1pt
\hfill {\tt hep-th/0503125}
\vskip0.2truecm
\begin{center}
\vskip 0.2truecm {\Large\bf
Cross sections for production of closed superstrings at high energy colliders 
in brane world  models
\vskip 0.2truecm
}

\vskip 2.5truecm

{\bf Diego Chialva $^{a,b}$
, Roberto Iengo $^{a,b}$ 
and Jorge G. Russo $^{c}$
}

\bigskip
\medskip


{{${}^a$
\it International School for Advanced Studies (SISSA)\\
Via Beirut 2-4, I-34013 Trieste, Italy} 

\medskip

{${}^b$ INFN, Sezione di Trieste}


\medskip

{${}^c$  
Instituci\' o Catalana de Recerca i Estudis Avan\c{c}ats (ICREA),\\
Departament ECM,
Facultat de F\'\i sica, Universitat de Barcelona,  Spain}
}
\bigskip

{\tt{chialva@sissa.it},
{iengo@he.sissa.it},
{jrusso@ecm.ub.es}
}

\end{center}
\begin{abstract}

In brane world string models with large extra dimensions, there
are processes where fermion and antifermion (or two gluons) can annihilate
producing a light particle (e.g. gluon) carrying transverse momentum 
and a Kaluza-Klein graviton or an excited closed string that
propagates in the extra dimensions.
In high energy colliders, this process gives a missing momentum signature.
We compute the total cross section for this process within the context
of type II superstring theory in the presence of a D brane. This
includes all missing energy sources for this string theory model up to
$s=8M_s^2$, and it can be used to put
new limits on the string scale  $M_s$.

\end{abstract}

\date{February 2004}

\vfill\eject

\tableofcontents


\setcounter{section}{0}


\section{Introduction}
\setcounter{equation}{0}

The existence of D branes has opened the possibility that our universe may have large 
(as compared to a fundamental or string length) extra dimensions
where only gravity propagates \cite{ADD}. Because the gravitational force is weak, the size of extra dimensions
might be as large as $O(0.1)$ mm \cite{hoyle}, for two large extra dimensions, and $O(10^{-11})$ mm, 
for six large extra dimensions.
Large extra dimensions imply that gravitational forces  can become strong
already at 1-10 TeV. This leads to  the exciting prospective that quantum gravity effects could be seen at LHC.

String theory provides a  theoretical framework to study the possible quantum gravity  physics at the TeV scale. 
An interesting process is the production of closed string states, 
gravitons but also other massless and massive states, that will propagate in the extra dimensions
and will appear as missing energy in the experiments \cite{antoniadis}.

Studies of  such processes where the closed string is
a Kaluza-Klein graviton  appeared in the literature
\cite{giudice,peskin0,dudas,peskin} (other effects of Kaluza-Klein
particles were investigated in \cite{hewett,bando,AAB,wells}).
Here we will focus on processes with quarks and gluons in the initial
states and an arbitrary (missing) closed string state and a gluon or a quark in the final state. 

We will follow the approach of \cite{peskin}:
in the ten-dimensional spacetime there are $N$ coincident D3 branes stretched out 
in three uncompact dimensions (our observed world), 
and the remaining 6 dimensions are compactified on a torus with periodicity $(2\pi R_m), \ m=1,...,6$.
The massless states associated with open strings that end on the D3 brane are described by
 ${\cal N}=4$ super Yang-Mills theory with gauge group $U(N)$ and 
the massless spectrum of open string states is thus given by the ${\cal N}=4$ supermultiplet.
Following \cite{peskin} we reinterpret its vectors and spinors as gluon and quarks.  

The present model \cite{peskin} is oversimplified,
but it is already capable of illustrating in a quantitative way the pattern that one should expect
to measure in a high energy collider due to missing energy processes,
when the CM (center of mass) energy overcomes the string scale.
We have made accurate calculations, 
taking care of factors $2$ and $\pi$, even though
we are aware that this is not a precision test of string theory.

We study the scattering of two gluons or of a quark and an anti-quark or of a gluon and a quark
giving as a final state a (missing) massless or massive closed string and a gluon or a quark
having a sizeable transverse momentum with respect to the incident beam. Hopefully 
the final state will experimentally appear as a detectable violation of transverse momentum conservation.

We consider the limit of transverse momentum $p_T$ small as compared to the string momentum scale 
$M_s=1/\sqrt{\a'}$
(but, given that
$M_s>O$(1) TeV, $p_T$ could be large as, say, 0.05 TeV or even more). 
In this limit we obtain
a very simple expression for the cross-section, in the form of the stringy cross-section for 
producing the closed string state from the initial state, as a function of the total CM energy only,
times a simple factor 
containing the dependence on $p_T$ and the emission angle $\theta$.
In a sense our case resembles a breehmsstralung process where the
``soft" emitted particle is the final gluon (or quark).

The string-theory cross-section for 
producing the closed string state from a initial state made of two
open strings can be obtained exactly from the imaginary part of
the forward scattering of two open strings at one non-planar loop.
This scattering amplitude is known in closed form in ten dimensions 
\cite{GS, schwarz}, and it can be easily generalized to the case of
open strings
on a Dp brane  \cite{peskin}.
 The cross section obtained in this way holds for any energy and
encodes all the allowed NS-NS and R-R (ten-dimensional) massless and massive closed string states, 
incorporating  also the Kaluza-Klein contribution.
This method is particularly useful in string theory, given the rapid
growth of the number of states at higher masses (for example, in type II
theory for level number $N=1$ there
are already  some thousands of propagating degrees of freedom).

Of course, in computing the imaginary part we 
do not meet any of the divergences, 
which instead can occur in the real part (the divergences in the real part
led to some puzzles in the evaluation 
of diagrams with virtual intermediate KK closed string states 
\cite{bando,dudas,peskin}). 

\smallskip

One distinctive feature of string theory, as opposed to other attempts to incorporate gravity in the theory, 
is the occurrence of an infinite 
number of unstable excited states, which appears as
resonances in the cross-section.

In the present case the resonances 
are due to open string intermediate states of mass 
$\a' M^2=N$, with odd $N$. These are essentially 
the poles that appear in scattering amplitudes in D branes
involving two open and
one closed string found by \cite{HK}.  
This characterizing signature of string theory 
could be seen in missing transverse momentum processes in colliders 
--and at the parton level it is particularly clean-- 
by looking at the dependence of the cross-section on $s$ 
(the total CM energy squared) either at fixed $p_T$ and $\theta$, 
or also after integrating over those variables in their allowed range.

We present our results for $0\leq\a' s\leq 8$, corresponding to the production of closed 
string states with (ten-dimensional) mass $\a' M^2=4N$ and $N=0,1$. 

One crucial ingredient to make quantitative 
estimates are the resonance widths that regularize the poles.
They are typically given by $\a_s\, M_s$ times some numerical coefficient,
where $\a_s$ is the strong coupling constant, $\a_s={g_{\rm YM}^2\ov 4\pi}$. 
We have estimated the widths
of the first resonances for various channels
(using the results of ref.\cite{peskin}) and used them  
as indicative average widths of the higher ones. \footnote{For a
  calculation of widths of closed string states, see  \cite{CIR}.}
We find two typical widths: $3.4 \a_s \, M_s$ and $7.4\a_s
\, M_s$
for gluon-gluon and quark-antiquark initial states respectively;
the comparison of the results for the two cases can give an idea on the dependence on the widths.  

Furthermore, some splitting of the resonance multiplets will occur, due in particular to
supersymmetry breaking. This will lead to
a splitting of some peaks in the cross section (this splitting will eventually smoothed out when including 
the effect of the parton distributions).

Plots of the parton level cross-sections for $gg\to g+${\it missing} and
$q\bar q\to g+${\it missing} are shown in section 5. 
We find  parton level cross-sections of order $25 pb\times  {1\ {\rm TeV}^2\over M_s^2}$ 
and $7 pb\times {1\ {\rm TeV}^2\over M_s^2}$ respectively, for the first peak.

\section{Total cross section for open+open to closed string}
\setcounter{equation}{0}

The particles in a high energy collider (electron-positron or quarks
and gluons) are represented 
in the string model by massless open string states. In this section
we are interested
in the total cross-section for two open massless string states going into 
an arbitrary closed string state, in the presence of a D3 brane.
In the absence of extra emission of a visible particle, this process
is not observed in colliders, since the closed string state escapes to
other dimensions and it cannot be detected.
However, we consider this process in detail since it will illustrate
the method and  some formulas will be used for the calculation of the missing-momentum
process of section 4.

\smallskip
Let us consider, for the beginning, the case of a D$_p$-brane in general. 
Take as $p_1^\mu,\ p_2^\nu$, $\mu,\nu =0,1,...,3$ the momentum of the
ingoing open strings and $P^{\hat \mu}$, $\hat \mu =0,1,...,9$, the
momentum of the outgoing closed string.
The center of mass energy is $s=2p_1\cdot p_2$, with  metric
$\eta_{\mu\nu}={\rm diag}(1,-1,-1,-1)$.
In the presence of the brane, translational invariance is broken in $9-p$ transverse directions and there is only
conservation of momentum along the $p$ directions of the brane.
 We choose the center of mass frame where $P_{\hat\mu }=(P_0,\vec
0, P_m)$, $m=p+1,...,9$.
In general, the formula for the cross section is as follows:
\be \label{generalcross}
\sigma_{2\to 1}(s)
&=&{1\over 2s} \int {d^9 P\over 2 (2\pi)^9 P_0}|A_{2\to 1}|^2 (2\pi)^p \delta^{(p)}(\vec
 p_1+\vec p_2-\vec P) 
(2\pi) \delta(E_1+E_2-\sqrt{M^2+\vec P^2})
\non\\
&=& {1\over 2s} \int {d^{9-p} P \over 2 (2\pi)^{8-p} P_0} |A_{2\to 1}|^2\delta(E_1+E_2-\sqrt{M^2+\vec P^2})
\non\\
&=&{\rm Vol}(
\Omega_{8-p})  {1\over 4s} (2\pi)^{p-8}|A_{2\to 1}(s)|^2 (s-M^2)^{{7-p\over
    2}}\ ,\ \ \ \ \ {\rm Vol}(\Omega_{8-p})= {2\pi ^{9-p\over 2}\over
  \Gamma( {9-p\over 2})}\ .
\label{crop}
\ee
Remarkably, the cross section does not have a peak at $s=M^2$. On the contrary, it vanishes.
The reason is that now $s=M^2$ is a threshold for the production of the state which in general will
have  $P_m$ transverse momentum components, and right on the threshold
$s=M^2$ the phase space vanishes.

\smallskip

The total cross section for the production of an arbitrary closed
string state from two open string states
can be obtained directly by taking the imaginary part of the amplitude corresponding to the non-planar annulus diagram
for forward scattering of $2\to 2$ open strings (with two vertex operators
 on the outer boundary, and two vertex operators
on the inner boundary). This amplitude
was obtained long ago by Green and Schwarz
\cite{GS, schwarz}. In the presence of a Dp-brane, we modify it by inserting, in the integrand over the modulus $\tau$, 
the factor $(\pi \tau ) ^{p-9\over 2}$ to take into account the 
non-conservation of the momentum transverse to the Dp-brane (see the analysis below and also \cite{peskin}):
\beq
A_{Dp} = {\cal N}_p\, K \, 2^{-2\a' s} \pi \int {d\tau } \ (\pi \tau ) ^{p-9\over 2} 
e^{{\pi \tau  \over 2} \a' s}
\int d\nu_{12} 
(\sin\pi \nu_{12})^{-\a' s} \int d\nu_{13} (\sin\pi \nu_{13})^{-\a' s} 
G(q,\nu_i)
\eeq
where 
\beq
G(q,\nu_i)=\prod_{n=1}^\infty { \big( 1-2 \cos(2\pi \nu_{13}) q^{2n-1}+
q^{4n-2}\big)^{\a' s}\big( 1-2 \cos(2\pi \nu_{24}) q^{2n-1}+
q^{4n-2}\big)^{\a' s}\over \big( 1-2 \cos(2\pi \nu_{12}) q^{2n}+
q^{4n}\big)^{\a' s}\big( 1-2 \cos(2\pi \nu_{13}) q^{2n}+
q^{4n}\big)^{\a' s}}\ ,\ \ \ q=e^{-\pi\tau }
\eeq
Setting $\xi_1=\xi_4$ , $\xi_2=\xi_3$, $t=0$, $s=-u$ (forward scattering), the kinematical
factor  for the case of two gluons
reduces to (see \cite{schwarz})
\beq
K=(\a' s)^2 \xi_2\cdot \xi_3 \ \xi_1\cdot \xi_4=
(\a' s)^2\, \xi_2^2 \ \xi_1^2\ ,
\label{kines}
\eeq
while for the case of two fermions (in the forward case $1\to 4,~ 2\to 3$)   
\beq
K=2(\a' s) \a' \bar u_1\gamma^{\mu}u_1 ~ \bar u_2\gamma_{\mu}u_2\ .
\label{kinef}
\eeq
As  we will take into account color quantum numbers  later on, here
we omit  Chan-Paton factors and, for a D3-brane, we take the same normalization
constant ${\cal N}_3$ as in \cite{peskin} 
\beq \label{norm}
{\cal N}_3= 8\alpha_s^2\pi^4 \qquad (p=3)\ .
\eeq
Consider the expansion in powers of $q$, 
\beq
G(q,\nu_i)=\sum_n c_n q^n=1+c_1q+c_2q^2+...
\eeq
The odd powers of $q$  drop out after integration in $\nu_i$. 
Upon integration in $\tau $, the $n$ power of $q$ produces a pole (for $p=9$) 
or a cut (for other $p$) at $\a' s=2n$, with $n=2N$, and
$N$ is a positive integer corresponding to the level of the string
excitation.
This reproduces the spectrum of the free closed superstring, $\a'
M^2=4N$. 
In conclusion, amplitude $A_{Dp}$ has cuts and
its imaginary part  picks  new contributions as
$s$ is increased from zero anytime it crosses $\a' s=4,8,12,...$ etc.
This is computed using the formula
\beq
{\rm Im}\int dx \ x^{-\a } e^{\beta x}={\pi \beta^{\alpha -1}\over 
\Gamma(\alpha )}\ .
\label{imas}
\eeq
This corresponds to having a closed string as intermediate state.
The reason of a cut, rather than a pole, is the presence of the Dp
brane. 
The phsyical process going on is as follows:
The two initial open strings meet into a single open string that
propagates
on the brane. The endpoints of this open string then meet and the
resulting open string goes into the bulk. Its transverse momentum is
absorbed by the brane. It is this intermediate state made of closed
string plus brane that produces the cut.
By using the optical theorem, we compute the
total cross section  for producing the closed string going into the bulk:
\beq
\sigma (s)= {1\over s} {\rm Im}A_{Dp}\ .
\eeq
The single open string propagating on the brane (just before the closed
string is produced) will give rise to poles
in the cross section. Such poles have first been observed by \cite{HK}.

Let us first compute the cross section up to $\a' s<4 $. In this
case, we can set $G=1$, since the remaining powers of $q$ give no
contribution to the imaginary part.
This will give  the level $N=0$ contribution, describing the production
of (Kaluza-Klein) gravitons.
Using the formula (\ref{imas}), we obtain
\be
{\rm Im}~A_{Dp} &=& {\cal N}_p\, K
{\pi  \big( { \a' s\over 2} \big)^{7-p\over 2}\over \Gamma ({9-p\over 2})} \left(
\int_0^1 d\nu (2\sin\pi \nu )^{-\a' s} \right)^2\ ,
\non\\
&=& {\cal N}_p\, K{\pi  \big( { \a' s\over 2} \big)^{7-p\over 2}\over \Gamma ({9-p\over 2})}\left( {\Gamma\big( {1\over 2}-
{\a 's\over 2}\big)2^{-\a' s}\over \sqrt{\pi } \Gamma  \big( 1-{\a
  's\over 2}\big)}\right)^2\ , \ \ \ \ \ s<8 \ .
\ee
Using the optical theorem, we have
 \beq
\sigma_0 = {\cal N}_p\, {1\over s} K
{  \big( { \a' s\over 2} \big)^{7-p\over 2}\over \Gamma ({9-p\over 2}) }
{2^{-2\a' s}\Gamma\big( {1\over 2}-
{\a 's\over 2}\big)^2\over  \Gamma  \big( 1-{\a 's\over 2}\big) ^2 }\ .
\label{afd}
\eeq
This can be compared with (\ref{crop}) when $M=0$. We see that the
volume factor ${\rm Vol}(\Omega_{8-p} )$ has emerged automatically, as
expected,
and also the power of $s^{7-p\over 2}$ coming from phase space.
It   agrees with
\cite{HK} where the amplitude for graviton emission from the brane was 
computed. 
The full cross section (\ref{afd}) includes --in addition to the
graviton-- the emission of other Neveu-Schwarz Neveu-Schwarz as well as
Ramond-Ramond states.

\smallskip

The  cross section (\ref{afd}) is valid up to $\a' s=4 $. As mentioned
before, for $\a' s>4$
there is a new contribution to the imaginary part due to the opening
of a new channel corresponding to $N=1$ level.
Including this contribution will give the exact expression for the
total cross section up to $\a' s=8$. For $\a' s>8$
there will appear a new contribution to the imaginary part
corresponding to $N=2$ level.
The coefficient of the quadratic $q^2$ term in $G$ is
\beq
c_2=2 \big( \a' s \cos(2\pi \nu_{12}) +1\big)  \big( \a' s \cos(2\pi \nu_{13})
+1\big)
+2(\a' s)^2\sin (2\pi \nu_{12})\sin (2\pi \nu_{13} )+ 2(\a' s )^2-2
\eeq
We will use the basic formula
\beq
\int_0^1 dx \big( \sin (\pi x) \big)^a \big( \cos(\pi x)\big)^{2b}
={1\over \pi } B({1\over 2} +{a\over 2},{1\over 2} +b)=
{\Gamma ({1\over 2} +{a\over 2}) \Gamma ({1\over 2} +b)\over\pi
\Gamma (1 +{a\over 2}+b) }\ .
\eeq
The final result for the cross section is
\beq \label{totalclosed}
\sigma= \sigma_0(s) + \hat \theta (\a's -4)~ \sigma_1(s )\ .
\eeq
where $\hat\theta (x) $ is the step function and
\beq \label{afd1}
\sigma_1(s)={\cal N}_p K \ {2\over  \Gamma ({9-p\over 2})} {1\over s}
\big( {\a' s\over 2} -2\big)^{7-p\over 2} {2^{-2\a ' s} \Gamma ({1\over 2}-{\a' s\over 2})^2\over\Gamma(1-{\a' s\over 2})^2}
B\ ,
\eeq
\beq
B \equiv \bigg({(\a' s )^2\over 2(1-{\a' s\over 2})}+1 \bigg)^2 + (\a'
s )^2-1\ .
\eeq

In conclusion, for the case of a D3-brane, we find the following
           formula for the total cross section 
(for an initial state with two gauge vectors):
\beq \label{crossschwarz0}
\s_0 (s) = {   \pi ^4       } (\xi_1^2\xi_2^2)  
 {\a_s^2\over M_s^2} \big( {s\over
  M_s^2} \big)^3\ {2^{-2\a ' s} \Gamma ({1\over 2}-{\a' s\over
           2})^2\over\Gamma(1-{\a' s\over 2})^2}\ ,
\eeq
\beq \label{crossschwarz1}
\s_1 (s)= 8    \pi^4 (\xi_1^2\xi_2^2)    {\a_s^2\over M_s^2}
 {s\over M_s^2} \big( {\a' s\over 2} -2\big)^2\ {2^{-2\a ' s} \Gamma
           ({1\over 2}-{\a' s\over 2})^2\over\Gamma(1-{\a' s\over
           2})^2} B\ .
\eeq

\section{Limits on cross sections and  widths of open string states}
\setcounter{equation}{0}

Consider a process $a\to b$, with a number of resonances in the
intermediate channel, with possibly
different widths
(as this is the typical case in string theory).
 
In our case, in particular, we have as first open string resonances: 
four scalars, one spin 1 
and one spin 2 resonances, denoted as $|0_{q=1,2,3,4}\rangle$, $|n_1\rangle$
and $|n_2\rangle$, respectively. They occur for $\alpha's=1$ (masses
$M_r=(\alpha')^{-{1\over 2}}$) and 
couple to various helicities (see \cite{peskin}).

The general cross-section formula for resonances of the same mass is
 \beq \label{sigmares}
 \sigma={16\ov s}{\pi \ov \mathcal{N}}\sum_j (2j+1)
  {M_r^2\Gamma_{ja}\Gamma_{jb} \ov (s-M_r^2)^2+\Gamma_j^2 M_r^2}\ ,
 \eeq
where we have indicated as $|a\rangle$ the initial states, as
 $|b\rangle$ the final ones (in our case, the closed string states).
The sum is over resonances with different
 spin $j$, whose total decay rate is $\Gamma_j$ ($\Gamma_{ja,b}$ are
 partial width).
$\mathcal{N}$ is the multiplicity of the initial states, that is
the number of possible polarizations times a possible Bose symmetry
factor. 

This formula can be obtained from unitarity (see appendix (A)),
generalizing a similar formula 
given in \cite{weinberg}.

In general $\Gamma_j=\Gamma_{ja}+\Gamma_{jb}+\Gamma_{j,other}$, so
that
 \beq \label{gammabound}
 {\Gamma_{j,a}\Gamma_{j,b}\ov \Gamma_j^2}\leq{1 \ov 4}\ ,
 \eeq
and the bound for $s=M_r^2$ is
 \beq \label{bound}
 \sigma\leq {4\pi \ov M_r^2\mathcal{N}}\sum_j (2j+1).
 \eeq
Considering the various initial states and the related resonances
(see below, in the text), we get the bounds:
$\langle\sigma_{gg}\rangle\leq {14\pi \ov M_r^2}$ for the gluon-gluon
initial state and $\langle\sigma_{q\bar q}\rangle\leq {18\pi \ov
  M_r^2}$ for the quark-antiquark one.

Using the results of \cite{peskin}, we can actually
compute the width of the resonances for $\alpha's=1$.

In the formula (\ref{crossschwarz0}) the width is put to zero. We can
compare it for $\alpha's \to 1$ with (\ref{sigmares}) when we
similarly consider $\Gamma_j \to 0$ in the denominator. Putting 
$\alpha' s = 1+y$, with $y\to 0$ we get
 \beq \label{resonancepole}
\alpha'\pi^4 \alpha_s^2 {1\ov \pi}{1\ov y^2}=\alpha'
	  {16\pi \ov \mathcal{N}}
          \sum_j (2j+1)\alpha'{\Gamma_{ja}\Gamma_{jb} \ov y^2}\ .
 \eeq

According to \cite{peskin} the various resonances couple to some
specific initial states (we consider QCD particles 
--~gluon $g$  and quark $q$~-- rather than QED
particles as in \cite{peskin}). Note that the resonances in our case are colourless.
 $$
 \mathbf{g_{R(L)}g_{R(L)}\to |0_1(0_2)\rangle }, \quad 
 \mathbf{\bar q_{R(L)}q^-_{R(L)} \to |0_3,(0_4)\rangle }, \quad
 \mathbf{g_L g_R \to |2\rangle }, \quad
 \mathbf{\bar q_Lq^-_R \to |1,2\rangle }
 $$

Take for instance $g_{R(L)}g_{R(L)}$: Here 
$\mathcal{N}_{gg}=1/2$ and using
$\Gamma_{01}(g_{L(R)}g_{L(R)})$ from \cite{peskin} we find
 \beq
 \sqrt{\alpha'}\Gamma_{01b}=\sqrt{\alpha'}\Gamma_{02b}=
  {\pi^2 \ov 8}\alpha_s\ .
 \eeq
Similarly we get
 \beq
 \sqrt{\alpha'}\Gamma_{03b}=\sqrt{\alpha'}\Gamma_{04b}={\pi^2 \ov
 8}\alpha_s\ ,
 \qquad
 \sqrt{\alpha'}\Gamma_{1b}={\pi^2 \ov 12} \alpha_s\ , \qquad
 \sqrt{\alpha'}\Gamma_{2b}={\pi^2 \ov 4}\alpha_s\ .
 \eeq

We now have to 
estimate the total widths, by summing over all possible
decay channels. In principle we should then include also weakly and
QED interacting particles, but we neglect this contribution since the
relevant coupling constant is one order of magnitude less then
$\alpha_s$. In this way we find:
 \be
 \Gamma_{01}=\Gamma_{02}=
   (2+{\pi^2 \ov 8})\alpha_s\, M_s \ , \quad & \quad
 \Gamma_{03}=\Gamma_{04}=
 (9+{\pi^2 \ov 8})\alpha_s \, M_s\ , \\
 \Gamma_1=({9\ov 2}+{\pi^2\ov 12})\alpha_s \, M_s \ ,
 \quad &  \quad
 \Gamma_2=({13\ov 10}+{\pi^2 \ov 4})\alpha_s \, M_s \ ,\ \ \ \  M_s={1\ov \sqrt{\alpha'}}\ .
 \ee

\paragraph{The cross-sections.}

In principle we should also know the widths of the higher open string
resonances ($\alpha's=3,5,7$). As an estimate we simply use the same
width for all of them, obtained by averaging over the various widths
for $\alpha's=1$, for the initial states $|gg\rangle $ or $|q\bar
q\rangle $.

Numerically we take ($\alpha_s=0.1$)
 \be
 \Gamma_{gg} & = & {(\Gamma_{01}+\Gamma_{02}+\Gamma_{2}) \ov 3}=
           {1 \ov 3}({53 \ov 10}+{\pi^2 \ov2}){\alpha_s \ov \sqrt{\alpha'}}=
             0.34 ~M_s\ ,\\
 \Gamma_{q\bar q} & = & {(\Gamma_{03}+\Gamma_{04}+\Gamma_{1}+\Gamma_{2})\ov 4}=
           {1 \ov 4}({119 \ov 5}+{7\pi^2 \ov 12}){\alpha_s \ov \sqrt{\alpha'}}= 
           0.74 ~M_s\  .
 \ee
These are the values we have used in plotting the cross-sections with
 formula (\ref{totalclosed}) with the substitution (\ref{reciperesonance}).
Note that $\Gamma_{q\bar q}$ is larger than $\Gamma_{gg}$ due to the sum over
generations and flavours.

A further comment: the bound (\ref{bound})
is independent of the normalization (\ref{norm}) and also our results will be less sensitive
to the the normalization than one could have expected.
In fact, a change in the normalization of the LHS
of (\ref{resonancepole}), will affect
the determination of $\Gamma_{jb}$ (in the numerator of (\ref{sigmares})),
and in turn the total width $\Gamma_j$ (in the denominator (of \ref{sigmares})).
Thus there will be some compensation, and  even more after integrating the cross-section
over an interval of the total energy.

\section{Cross section for  missing energy processes in colliders}
\setcounter{equation}{0}

In \cite{giudice}, the cross section for the process of Kaluza-Klein  graviton $G$ emission in electron-positron
annihilation, $e^+e^-\to \gamma G$,  was  determined.
This  process gives a missing-energy signature, which has been used by
LEP2 experiments \cite{LEP2} to put constraints on the size of extra
dimensions. This process was further studied in \cite{peskin0}, in particular,
in \cite{dudas,peskin},  where the string theory corrections  were computed.

In general, a process gluon+gluon (or fermion+antifermion) $\to $ gluon
+ missing transverse momentum can have many  contributions of
different origin.
In  string theory, in addition to the graviton
considered in \cite{dudas,peskin}, there are also 
states from the Neveu-Schwarz Neveu-Schwarz and
Ramond-Ramond sectors which contribute to the cross section. 
In particular, there are massless NS-NS and  RR states which in four dimensions are scalar or vector 
particles. They  are expected to get a mass because a potential is generated 
for the scalars and the U(1) gauge symmetry is also broken by mechanisms 
which are strongly model dependent, involving fluxes, compactification scales 
and details of supersymmetry breaking. 
Their expected masses are also  model dependent. 
Here we treat them as massless, which is justified if their masses are much smaller than the string scale.

Here we will compute the cross-section which will
include, in addition to the process of Kaluza-Klein
graviton emission studied in earlier works, 
emission of RR states and emission of closed string
states of the first excited.
The contribution from the first excited closed string states
appears  for $\a' s>4$.

Thus the total cross-section obtained here represents the full cross
section for missing energy processes that takes into account all
possible string state production up to $s< 8M_s^2$. 
These results can be used to put new constraints on the string scale.

\smallskip

We consider the amplitude 
\beq
A_C=A(a+b\to C+c)
\eeq
where $a,b,g$ are gluons or quarks and $C$ is a (missing) closed string state.
The closed string state $C$ has momentum 
$p_{\hat \mu}=(E_C,{\vec p}_C,\Delta_i)$, $\hat \mu =0,...,9$, $i=4,...,9$,
with $\Delta_i$ being the components transverse
to the brane. Its mass in ten dimensions is
$M_C^2=E_C^2-{\vec p}_C{}^2-\Delta^2$. This corresponds to  a four dimensional mass
equal to $M^2_4\equiv E_C^2-{\vec p}_C{}^2=M_C^2+\Delta^2$. Since the six extra dimensions are
compactified on a 6-torus, $\Delta_i$ are quantized in units of inverse
radius of the toroidal directions. When the extra dimensions are very large compared to the
TeV scale we can consider $\Delta_i$ as a continuous variable and for simplicity we make this
assumption here. We have also considered  the case of two large and and four
small extra dimensions, and obtained very similar results, reported in
Sect. \ref{results}.

{} For center of mass energies $\sqrt{s}< 2 M_s$, the particle $C$ can
only be a massless closed string state. The cross section picks an
important extra contribution  when  $\sqrt{s}> 2 M_s$ coming from
all bosonic closed string states of the $N=1$ level, with mass $ M=2 M_s$.
The final cross section will be a function of $s$, $M_C$, the missing
energy $E_C$ {\it and} of the scattering angle $\theta $.
It is important to note that in ordinary
  four dimensional processes, 
the four-point amplitudes
depend on two kinematical variables, e.g. $s$ and $t$, or $s$ and
$\theta $. The reason of the dependence on  $s$, $E_C$ and $\theta $
in the present case is due to the momentum non-conservation in the
directions transverse to the brane.

We will consider as initial states in the collision: i) two gluons, ii) quark and antiquark,
iii) quark and gluon (an initial state of two quarks cannot produce a colorless closed string state).
In cases i,ii) the final particle c is a gluon, in the case iii) it is a quark.

We begin by illustrating the general computation taking the case i).

We  consider the CM frame, with  momenta and polarization tensors as follows:
\beq
p^a_\mu=(p_a,\vec p_a)\ ,\ \ \ p^b_\mu=(p_a,-\vec p_a)\ ,\ \  \ \
p^a_\mu p^{a\mu}=p^b_\mu p^{b\mu} = 0\ ,
\eeq
\beq
p^c_\mu=(p_c,\vec p_c)\ ,\ \ \ P^C_{\hat \mu}=(E_C,-\vec p_c,\Delta_i)\ ,\ \  \ \
p^c_\mu p^{c\mu}=0\ ,\ \ 
E_C^2-\vec p_c{}^2-\Delta^2=M_C^2\ ,
\eeq
\beq
E_C+p_c=2p_a\ ,
\eeq
\beq
\vec \zeta^i\cdot \vec p\, {}^i=0\ ,\ \ \ \zeta^i_0=0\ ,\ \ \ \ i=a,b,c\ .
\eeq
We  consider
the case in which the gluon $c$ is {\it soft}, i.e. it carries an
energy much less than the string scale.
More precisely, we assume $p_c << {1\ov \a 'p_a}$.
In the CM frame, $s=4 p_a^2$.
Since the string theory corrections are significant for $s>O(M_s)$ and
$M_s>O(1\ TeV)$, the gluon $c$ could even carry energies as large as,
say, 0.1 $TeV$, and the present approximation will miss only
corrections of few per cent. Given the breehmsstralung character of
the emission process, the approximation of a soft emitted gluon will
be accurate  for most missing-energy events that will appear in a collider
(this is confirmed by the cross section obtained below, which decreases like $1/p_c$).


We will call, then, $V_{C}$ the vertex (to be used in Feynman graphs) 
for producing 
the state $C$ from two (real or virtual) gluons $i$ and $j$:
\beq
\label{Cprodvertex}
V_C=\zeta^i_{\mu}\zeta^j_{\nu} T^{\mu\nu A}\epsilon_{A}\ ,
\eeq
where $\zeta^{i,j}$ are the gluon polarization and $\epsilon$ is
the polarization of the closed string state equipped with whichever indices
$A$ are necessary.

Note that
\beq
\sum_{\epsilon}\int|V_C|^2 {d\Omega_5 \Delta^4\ov 2(2\pi )^5} = 2 {\rm
  Im}[A_{D3}]=2\hat s~
(\sigma_0(\hat s)+\theta (\hat s-4)\sigma_1(\hat s) )\ .
\label{nhy}
\eeq
Here $A_{D3}$ is eq.(2.2) for the case $p=3$,
$\vec\Delta$ is the closed state KK-momentum transverse to the brane
and $\hat s=M^2_C+\Delta^2=s-2p_c\sqrt{s}$ is the square mass including the KK-momentum.

\vskip0.5cm

The cross-section for the process $a+b \to C+c$ will be
\beq
d\sigma={1\ov 2s}\int |A_C|^2{d^6\Delta d^3p_C\ov 2(2\pi )^9E_C}{d^3p_c\ov 2(2\pi )^3p_c}
(2\pi )^3\delta^3 (\vec p_C+\vec p_c)(2\pi )\delta (2p_a-p_c-E_C)=
{1\ov 2s}F{d^3p_c\ov 2(2\pi )^3p_c}
\eeq
where 
\beq
F=\int |A_C|^2{d^6\Delta \ov 2(2\pi )^6 E_C}(2\pi )\delta (2p_a-p_c-E_C)=
\int |A_C|^2 {d\Omega_5\Delta^4 \ov 2(2\pi )^5} \ .
\label{F}
\eeq

By dimensional reasons $[F]=p^{-2}$ and we look for the leading terms
that is the ones proportional to $p_c^{-2}$ since $p_c$ is the smallest energy scale. 
These terms can only come from the (modulus square of the sum of the) Feynman graphs having a pole
in the Mandelstam variables $t=-2p_a\cdot p_c,u=-2p_b\cdot p_c$, due to an
the intermediate virtual gluon.
Therefore we compute $A_C$ by considering only those two graphs (see
fig.~1, I and II).

\begin{figure}[ht!]\label{fgraph}
\centering
\includegraphics*[width=250pt, height=120pt]{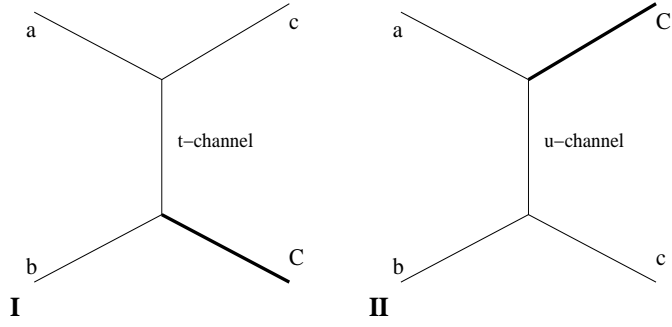}
\caption{$t$- and $u$-channels for the process: 
 a+b $\to$ c+C (missing, closed-string state).}
\end{figure}

The resulting cross-section will be of the form
\beq 
d\sigma={1\ov s}f d\Omega_2{dp_c\ov p_c} \ ,
\eeq
where $f=f(s,\hat s,\a ' )$
is dimensionless. Notice the factor ${dp_c\ov p_c}$ characteristic of a
breehmsstralung process, in this case the breehmstralung of the gluon $c$
which is the {\it soft} particle in our process.

\vskip0.5cm

Take for instance the $t$-channel graph (see fig. \ref{fgraph}, I)
\beq
V_{a\to c+e}\cdot {\it Propagator}_{e}\cdot V_{C}
\eeq
where 
${\it Propagator}_{e}\sim{1\ov p_e^2}$ and $V_C$ is the vertex for $b+e\to
C$ defined in (\ref{Cprodvertex}). 

\vskip0.5cm

The polarization of $e$ will be computed in terms of the polarizations and 
the momenta of $a$ and $c$ by the YM vertex $V({\rm YM})_{a\to c+e}$.
One can check that the resulting polarization satisfies $\zeta_e\cdot p_e=0$.

We get:
\beq
A_C=-{\zeta_{b\mu}\zeta_{e\nu} T^{\mu\nu A}\epsilon_{A}\ov 2p_ap_c}\ .
\eeq

Note that 
$|p_e^2|=|(p_a-p_c)^2|=2p_ap_c << {1\ov \a '}$  and therefore, since the scale is set here by $1/\alpha '$,  
we can take $V_{C}$ to be the vertex 
to be computed in string theory for massless $b$ and $e$.
Thus eq.(\ref{nhy}) holds.

By inserting $|A_C|^2$ in (\ref{F}) we get
\beq
F=W\cdot 2 {\rm Im}[A_{D3}]\ ,
\eeq
where $W$ depends on $p_c$ and $\theta$ and will be explicitly computed below for 
the various cases i,ii,iii).

Therefore 
\beq
d\sigma_C={\hat s\ov s}~
(\sigma_0+\hat \theta (\hat s-4)\sigma_1 )W{d^3p_c\ov 2(2\pi )^3p_c}\ .
\eeq
In our approximation $\alpha' (s-\hat s)<< 1$ and thus we can take $\sigma_{0,1}$ to depend on $s$.
Therefore the only dependence on $p_c,\theta$ is inside $W$.

\vskip1cm

\noindent {\bf i) gluon-gluon initial state.} We recall the
kinematical factor $K$ (\ref{kines})
\beq
K=(\alpha' s)^2(\zeta^1\cdot\zeta^4)( \zeta^2\cdot\zeta^3)  \ ,
\eeq
for the amplitude
$A_{D3}=g+g\to {\rm closed\ string}\to g+g$ (forward scattering
corresponds to $1\to 4$, $2\to 3$).

We use $\zeta ^a=\zeta_{\mu s} ^a$ to indicate the particle $a$ with color $s$ etc.
For the (t,u)-channel we get
\beq
\zeta^e_{e\rho} =\zeta^{a,b}_{\mu s}\zeta^c_{\nu r}V({\rm YM})^{\nu\mu\rho}_{sre}
=g_{\rm YM}f_{sre}(\zeta^{a,b}_s\cdot\zeta^c_r(p_{a,b}+p_c)_{\rho}-2p_c\cdot\zeta^{a,b}_s\zeta^c_{\rho r}
-2p_{a,b}\cdot\zeta^c_r\zeta^{a,b}_{\rho s})\ .
\eeq
We keep only the last term of this expression since the others give sub-leading contributions (see the 
contractions in the following formulas). Thus we define:
\beq
\zeta^{t,u}_{\mu e}= -g_{\rm YM}(2p_{a,b}\cdot\zeta^c_r)\hat\zeta^{t,u}_{\mu e}
\ \ \ ({\zeta_{\mu s}^{a,b} f_{sre}}\equiv\hat\zeta^{t,u}_{\mu e}) \ , 
\eeq
\beq
t,u=p_{(t,u)}^2 \ ,\ \ \ \ p_{(t,u)\mu}=p_{(a,b)\mu}-p_{c\mu} \ .
\eeq
Note that in the vertex producing the closed string state, the color indices of the two gluons
are contracted together, since the closed state is colorless (technically 
this is due to the Chan-Paton rules \cite{peskin}).   
We have  (since $p_b\cdot\zeta ^c=-p_a\cdot\zeta ^c$) 
\beq
\sum_{\epsilon}\int {d\Omega_5 \Delta^4\ov 2(2\pi )^5} 
g_{YM}^24(p_a\cdot\zeta^c )^2|({\hat\zeta^{t}_{\mu e}\zeta_{\nu s} ^b \delta_{es}\ov t}-
                             {\hat\zeta^{u}_{\mu e}\zeta_{\nu s}^a
			       \delta_{es}\ov u})T^{\mu\nu
			       A}\epsilon_{A}|^2 = 
{2 {\rm Im}[A_{D3}]}
\eeq
where in ${\rm Im}[A_{D3}]$ we have to replace $(\zeta_1\cdot\zeta_4)(\zeta_2\cdot\zeta_3)$  with  the term $W$
\beq
W= g_{YM}^24(p_a\cdot\zeta^c_r )^2\left({( \zeta^a_s\cdot\zeta^a_{s'}) ( \zeta^b_h\cdot\zeta^b_{h'}) \ov t^2}
+{(\zeta^b_s\cdot\zeta^b_{s'})(  \zeta^a_h\cdot\zeta^a_{h'}) \ov u^2} 
 - 2{(\zeta^a_s\cdot\zeta^a_{h'})(\zeta^b_h\cdot\zeta^b_{s'})\ov tu}\right) f_{srh}f_{s'rh'}  \nonumber
\eeq
here $ \zeta^b_s\cdot\zeta^b_{s'} =\zeta^b_{\mu s}\zeta^b_{\mu s'} $ and 
the sum is over colors $s,s',h,h'$ at fixed $c$.

The last interference term corresponds to the  amplitude
(2.2) with 
$\zeta^t+\zeta^b \to \zeta^a+\zeta^u$  which is forward in the sense that
$(p^t-p^a)^2=0, \  (p^b-p^u)^2=0$.

We average over $\zeta^a,\zeta^b$ and sum over $\zeta^c$ getting for SU(N)
\beq
W =   {g_{\rm YM}^2 N\ov (N^2-1)}\, {1\ov p_c^2}\, {4\ov \sin^2(\theta
  )}\ .
\eeq
In conclusion we get\footnote{We have checked that our result
coincides with the expression obtained using the amplitude 
in eq.(4.14) of \cite{dudas}, 
in the particular case considered there, by taking the relevant limit and normalization.} 
for SU(3), after integrating over $\phi$,
\beq \label{ggCc}
d\sigma=
(\sigma_0+\hat \theta (\a' s-4)\sigma_1 )\alpha_s {2\ov \pi}{3\ov  8}
{dp_c\ov p_c} {d\theta \ov \sin(\theta)}\ .
\eeq

\bigskip

\noindent {\bf ii)  quark-antiquark initial state.} We recall the
kinematical factor $K$ (\ref{kinef})
\beq
K=\alpha' s 2\alpha' (\bar v_{2s}\gamma^{\mu}v_{3r})(\bar
u_{4r}\gamma_{\mu}u_{1s})\ ,
\eeq
for the amplitude
$A_{D3}=q+\bar q\to {\rm closed\ string}\to q+\bar q$, 
that is $u_1(k_1)+\bar v_2(k_2)\to u_4(k_4)+\bar v_3(k_3)$ (forward is $1\to 4$ and $2\to 3$).
We have indicated the color indices $s,r$. We normalize the spinors by requiring that
the current is the same as in the bosonic case: $\bar u\gamma^{\mu}u=p^{\mu}$.
With this normalization the Dirac average is  $\langle 
u_{s}\bar u_{r}\rangle ={p\cdot\gamma\delta_{sr}\ov 4}$.

Our initial state is $u^a, \bar v^b$ and we call $u^t$ ($\bar v^u$) the intermediate quark (anti-quark).
Let us call $\bar v V^Au \epsilon_A$ the vertex for producing the closed state from 
$q+\bar q$. The leading amplitude for $F$ comes from the two graphs with
a pole in the $t$ and $u$ channels
\beq
A_t=\bar v^bV^A u^t \epsilon_A \ ,\ \ \ \ \ 
 A_u=\bar v^u V^Au^a \epsilon_A\ .
\eeq
Here 
\be
u^t &=& g_{YM}{(p_{t}\cdot\gamma)(\zeta_c\cdot\gamma) t_c u^a\ov t} ~~
{\rm with} ~~~ p_t=p_a-p_c \ ,\nonumber \\
\bar v^u &=& -g_{YM}{\bar v^b t_c \zeta_c\cdot\gamma p_u\cdot\gamma\ov
  u} ~~ 
{\rm with} ~~~ p_u=p_b-p_c\ .
\ee
 $t_c$ is a SU(N) matrix in the fundamental representation $Tr(t^2_c)=1/2$.
Now we have
\beq
\sum_{\epsilon}\int {d\Omega_5 \Delta^4\ov 2(2\pi )^5} |A_t+A_u|^2 =
2{\rm Im}[A_{D3}]\ ,
\eeq
where in ${\rm Im}[A_{D3}]$ we have to replace
 $(\bar u_{2s}\gamma^{\mu}u_{3r})(\bar u_{4r}\gamma_{\mu}u_{1s})$  with  the term
\beq
Q=(\bar v^b_s\gamma^{\mu}v^b_r)(\bar u^t_r\gamma_{\mu} u^t_s)+
       (\bar u^a_s\gamma^{\mu}u^a_r)(\bar v^u_r\gamma_{\mu} v^u_s)+
       2(\bar v^b_s\gamma^{\mu}v^u_r)(\bar u^a_r\gamma_{\mu} u^t_s)\ .
\eeq
The last interference term corresponds to the amplitude (2.2) with $u^t (p_t)\to u^a(p_a)~~v^b(p_b)\to v^u(p_u)$,
which is forward: $\hat t =(p_t-p_a)^2 =(p_b-p_u)^2=p_c^2=0$.

Now we substitute the expressions for $u_t, v_u$ and we make the Dirac
 and color average over our initial states. We find:
 \beq
 \langle (\bar v^b_{s}\gamma^{\mu}v^b_{r})(\bar u^t_{r}\gamma_{\mu} u^t_{s})\rangle = 
{g^2\ov t^2}{Tr(t_c^2)\ov N^2}p_2^{\mu}
\langle \bar u^{a}\zeta_c\cdot\gamma p_t\cdot\gamma\gamma_{\mu}
p_t\cdot\gamma\zeta_c\cdot\gamma u^a\rangle \to
{g^2\ov N^2 8t^2}{s\ov 2}16(p_1\cdot\zeta_c )^2\ .
\eeq
We have kept only the leading terms (those proportional to four powers of $p_a=p_b$) 
noting that $p_t\cdot\zeta_c \sim p_a\cdot\zeta_c ,~~ p_b\cdot p_t \sim p_a\cdot p_b$.
Similarly
\beq
\langle (\bar u^a_{s}\gamma^{\mu}u^a_{r})(\bar v^u_{r}\gamma_{\mu}
 v^u_{s})\rangle \to
{g^2\ov N^2 8u^2}{s\ov 2}16(p_1\cdot\zeta_c )^2\ .
\eeq
For the last term we get
\beq
  \langle \bar v^a_{s}\gamma^{\mu}v^u_{r}\rangle \langle \bar u^a_{r}\gamma_{\mu} u^t_{s}\rangle 
\to  {g^2\ov N^2 32tu}{s\ov 2}64(p_1\cdot\zeta_c )^2\ .
\eeq
The final result is
\be
\alpha' s 2\alpha' Q &=& (\alpha' s)^2  2{g^2\ov N^2}(p_1\cdot\zeta_c )^2({1\ov t^2}+{1\ov u^2}+{2\ov tu}) \nonumber \\
&=& (\alpha' s)^2 2{g^2\ov N^2} {(p_1\cdot\zeta_c )^2\ov
  4p_1^2p_c^2}{4\ov (1-\cos^2(\theta ))^2}\ .
\ee
We can express the result of the quark-antiquark case by saying that:
in ${\rm Im}[A_{D3}]$ we have to replace 
$(\zeta_1\cdot\zeta_4)(\zeta_2\cdot\zeta_3)$ (case of the gluons)
with  the term $W$, which after summing over the polarizations and
the colors of the emitted gluon is
\beq
W={g^2(N^2-1)\ov N^2} {2\ov p_c^2 \sin^2{\theta}}\ .
\eeq
In conclusion, after integrating over $\phi$, in the case of SU(3) we get
\beq \label{qqCc}
d\sigma=
(\sigma_0+\hat \theta (\a' s-4)\sigma_1 )\alpha_s {8\ov 9\pi}{dp_c\ov p_c} 
{d\theta \ov \sin{\theta}}\ .
\eeq

\bigskip

\noindent {\bf iii) quark-gluon initial state.} We find that in this case $d\sigma$ is sub-leading with respect to i) and ii) by a factor $p_c/p_a$.
This is due to the fact that in the t-channel we get a gluon $\zeta^t_{\mu} \sim \bar u(p_c)\gamma_{\mu}u(p_b)$
and in the u-channel we get a fermion $u^u \sim \gamma\cdot p_u\gamma\cdot\zeta^a u(p_c)$. Thus in the average
over $u(p_c)$ there is an extra power of $p_c$  in the numerator.

\vskip0.5cm

{\bf Integrated cross-section.} The result for $d\sigma$ are the same, up to a constant which is also numerically rather
similar, for the cases {\bf i)} and {\bf ii)}, see eqs.(\ref{ggCc},
\ref{qqCc}).
In our approximation $\sigma_{0,1}$ are independent of $p_c$ and $\theta$. 
Therefore we can easily perform the integration on those variables, in particular for a range $\Delta\theta$ around $\pi /2$ 
such that $sin(\theta )\sim 1$. We get
\beq \label{missingcross}
\sigma=\mathcal{A} {\alpha_s\ov\pi}\big(\sigma_0+\hat \theta (\a's-4)\sigma_1 \big)\,
\log {p^{max}_c \ov p^{min}_c}\ \Delta\theta\ ,
\eeq
where $\mathcal{A}$ is $6/8$ for case {\bf i)} and $8/9$  for case {\bf ii)}. 

\section{Analysis of the parton cross sections}\label{results}                                          
\setcounter{equation}{0}

In the previous Section we have evaluated the cross section for the case in which the final state contains, besides the 
``missing'' closed string, a gluon with  transverse momentum (which is
small as compared to $M_s$), 
as a function of the center of
mass energy squared up to $s=8M_s^2$.
The cross section is proportional to (\ref{totalclosed}), (\ref{crossschwarz0}), (\ref{crossschwarz1}) and has poles coming from the Gamma functions at $\a' s
=1,3,5,7$, representing open string resonances. The
 poles have been regularized by including  finite widths of the open string
resonances (of order
$\Gamma=O(\a_s M_s)$) according to the following recipe.
We re-write the factor $\Gamma({1 \ov 2}-{\alpha's \ov 2})$ in
(\ref{crossschwarz0}), (\ref{crossschwarz1}), which contains the poles,
as:
 \beq \label{reciperesonance}
 |\Gamma({1 \ov 2}-{\alpha's \ov 2})|^2=
{\pi^2 \ov \Gamma({1\ov 2}+{\alpha's \ov 2})^2}{1\ov \cos^2(\pi{\alpha's\ov 2})}
 \to
{\pi^2 \ov \Gamma({1 \ov 2}+{\alpha's \ov 2})^2}{c(\Gamma)\ov
  |\cos\big( {\pi\over 2}(\alpha's+i\alpha'\sqrt{s}\Gamma )\big)|^2}\ ,
 \eeq
with $c(\Gamma)={4 \ov \alpha'\Gamma^2\pi^2}~ \sinh^2{({\pi \ov 2}\sqrt{\alpha'}\Gamma)}$ 
in order to match the l.h.s
of (\ref{resonancepole}) when $y^2 \to y^2+\alpha'\Gamma^2$ for 
$y \to 0$. 

The widths of open string resonances $\a' s=2n+1$, with $n\geq 1$, are 
not known.
We have taken in (\ref{reciperesonance})
the numerical value of $\Gamma$ at $n=0$ estimated in Section 3,
considering for all the different resonances the same
averaged value (either $\Gamma_{gg}$ or $\Gamma_{q\bar q}$) depending
on the initial state, consisting of two gluons ($gg$) or fermion and
anti-fermion ($q\bar q$). 
One expects  more massive resonances to have larger widths since at higher level there is more phase space
and more decay channels available. This fact is qualitatively taken into account by our prescription (\ref{reciperesonance}),
which near a pole $\a' s=2n+1$ reads
\beq 
|\Gamma({1 \ov 2}-{\alpha's \ov 2})|^2
 \to {4\ov n!\, \a' \Gamma^2}
\left({\sinh{({\pi \ov 2}\sqrt{\alpha'}\Gamma)}\ov \sinh{({\pi \ov 2}\sqrt{2n+1}\sqrt{\alpha'}\Gamma)}}\right)^2
 \eeq
so that the width of the resonance effectively increases as $n$ is increased.

We have seen that the differential cross sections for gluon-gluon eq.(4.21) and quark-antiquark eq.(4.32) are proportional to 
$\sigma =\sigma_0+\hat \theta (\a's-4)\sigma_1$, the difference being the numerical factor in front and, more importantly, the widths.
In Fig.2 we show 
$p_c \sin\theta \times {d\sigma/ dp_cd\theta}$
for both cases, in units of $\a'$, as a function of
$\a' s$ ($s$ being the square of the CM energy). 
Note that, in the range of validity of our approximations,
this is independent of $p_c$ and $\theta $.
We take $\alpha_s =0.1$.
 
As explained in the previous section, the case gluon-quark $\to$ closed-string(missing)+quark is sub-leading
and quark-quark $\to$ closed-string(missing)+any is forbidden by color symmetry.

\begin{figure}[ht!]\label{gluons}
\centering
\includegraphics*[width=250pt, height=180pt]{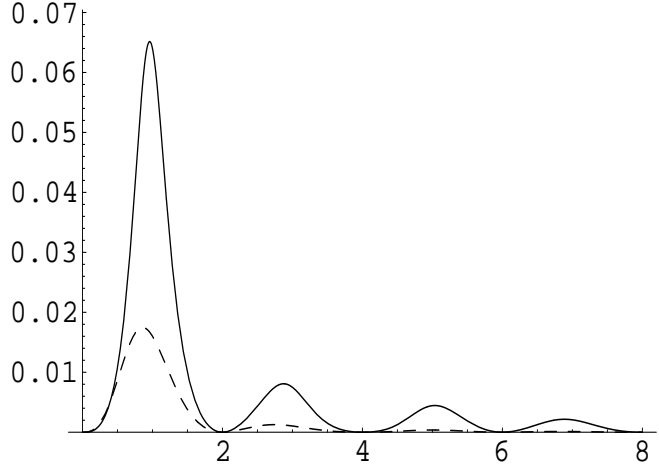}
\caption{Process gluon+gluon $\to$ closed-string(missing)+gluon (continuous line) and 
quark+anti-quark $\to$ closed-string(missing)+gluon (dashed line), 
cross-section in units of $\a'$, as a function of  $\a' s$.}
\end{figure}

\begin{figure}[ht!]\label{compact}
\centering
\includegraphics*[width=250pt, height=180pt]{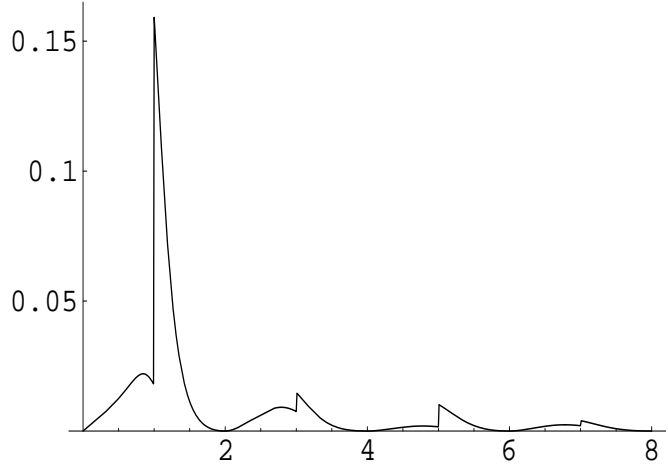}
\caption{Process gluon+gluon $\to$ closed-string(missing)+gluon in the case of two large  
and four small (with radii $R=\sqrt{\a'}$) extra dimensions.}
\end{figure}

\smallskip
\noindent In Fig.3 we show the gluon-gluon cross section in the case when two of the extra dimensions 
are large and the remaining four dimensions are small with radius $R=\sqrt{\a'}$.
The relevant cross-section formula is obtained
by replacing,  in the phase space factor in (\ref{generalcross}),
the integrals over momenta in the small
compactified dimensions by the sums over the Kaluza-Klein modes. 
This in turns amounts to simply substitute in
formula (\ref{missingcross})
 \beq
 (s-M^2)^2\hat \theta (s-M^2) \to {2 \ov \pi^2 R^4}
 \sum_{n_1,n_2,n_3,n_4} \hat \theta(s-\sum_{j=1}^4 {n_j^2
   \ov R^2}-M^2)\ .
 \eeq  
which produces a number of discontinuities in the cross section. 

In order to make predictions for the measurable $pp$ cross section which is relevant for LHC, 
the parton cross sections given here have to be convoluted with the parton distribution functions.
In addition, one would need to study the dependence on physical observables for jets.
The main effect of the parton distribution functions will be to smooth out the various structures 
of the parton cross sections, like peaks and discontinuities. It would be interesting to 
investigate whether the possible effects of the string resonances could be recognized from the data.
This requires a separate analysis by collider phenomenology experts.

\section{Concluding remarks}
\setcounter{equation}{0}

Missing energy processes provide an efficient way  to look for large extra dimensions.
In brane world string models, there are many possible processes
leading to missing
energy, coming from the large number of string excitations.
Here we have described a simple method to incorporate in a closed form all 
string theory contributions
to processes of missing energy  by expressing
the  cross section of the process $gg\to g+$missing (or  
$q\bar q\to g+$missing) in terms 
of the imaginary part of a  non-planar
amplitude. We have also made an estimate of the decay widths.

The cross section at the peaks is very sensitive to the decay widths.
In realistic models the widths could be smaller than
those presented here, because after supersymmetry breaking some of the $N=0$ states
become massive and the phase space for decaying into them is reduced.
As a result, the size of the cross section could be bigger.

Another interesting feature of the total cross section is its smooth
behavior after going through the threshold for production of closed string states of
level number $N=1$.
We have checked that for $\a' s>4$ the dominant contribution comes from the
$N=1$ states, while the contribution of the $N=0$ states is significantly
smaller.
In general, the string-theory cross section is of the form  
$$
d\sigma (s)= \sum_{N=0}^\infty \ \sigma_N(s) \ \hat \theta(\a' s-4N)\ .
$$
In this paper we have computed two terms  $\sigma_0(s)$ and   $\sigma_1(s)$,
which are sufficient in order to explore CM energies up to $s=8M_s^2$.
All remaining terms give vanishing contribution at  $s<8M_s^2$.
It is worth noting that for large $s$ both terms $\sigma_0(s)$ and
$\sigma_1(s)$  go exponentially
to zero, due to the factor $ 2^{-2 \alpha' s}$ in
(\ref{crossschwarz0}), 
(\ref{crossschwarz1}).
It is plausible that this will be true for all $\sigma_N$, i.e. that
each individual term (corresponding to all string contributions of a given
mass $M=2\sqrt{N}M_s$)
 is exponentially suppressed at large $s$.
The explicit form of $\sigma_0$ and $\sigma_1$ suggests that at large
$s$ a
generic term should have the form 
$\sigma_N(s)\cong c_N \a' \alpha_s^2\, (\a' s)^{2N}\, (\a's-4N)^2\ 2^{-2 \a' s}$.
It would be interesting to see if this is indeed the case.

In conclusion,
the cross section for the leading missing transverse momentum process
exhibits a typical pattern with peaks and zeroes at periodic values
of the CM energy, where the widths and the heights of the peaks are given
in terms of the string scale $M_s$ and $\alpha_s$. 
We provide a quantitative picture of this pattern, taking a basic string model,
see figs.(2, 3).  
We hope that the present results
may help to look for
experimental signals of genuine string theory effects at high energy
colliders.

\section{Acknowledgements}
The authors acknowledge partial support by the EC-RTN network
MRTN-CT-2004-005104. D.C. and R.I. also acknowledge partial support 
by the Italian MIUR program ``Teoria dei Campi, Superstringhe e Gravit\`a".
J.R.  also acknowledges   partial support by MCYT FPA
2004-04582-C02-01 and CIRIT GC 2001SGR-00065.


\setcounter{section}{0}
\appendix{Cross-sections in presence of resonances} \label{resonances}

The formula (\ref{sigmares}) can be obtained from unitarity, as follows.

\smallskip

Near  resonances, we expect the S-matrix to have the following form:
 \beq
 S=S_0+{R_1 \ov E-E_r+i{\Gamma_1\ov 2}}+{R_2 \ov E-E_r+i{\Gamma_2\ov 2}}
 \eeq
where we have supposed that there are two resonances at the same
energy with two different decay widths.

By unitarity:
 \beq
 S^\dagger S=1\ ,
 \eeq
whence
 \beq
 \begin{cases}
 S_0^\dagger S_0=1 &  \\
 (E-E_r+i{\Gamma_i\ov 2})R^\dagger_iS_0+(E-E_r-i{\Gamma_i\ov
 2})S_0^\dagger R_i+R^\dagger_iR_i=0 &  \\
 R^\dagger_1R_2+R^\dagger_2R_1=0 &
 \end{cases}
 \eeq

We can parametrize
 \beq
 R_i=-i\Gamma_iA_iS_0\ ,
 \eeq
and the system becomes:
\beq
 S_0^\dagger S_0=1\,;  \qquad
 A^\dagger_i=A_i\,;   \qquad
 A^2_i=A_i\,; \qquad
 A^\dagger_1A_2+A^\dagger_2A_1=0
\eeq

 Therefore we can write
 \beq
 A^i=A^i_{out, in}=\sum_{r_i} u_{out}^{(r_i)}u_{in}^{(r_i)*}
 \eeq
where $r_i$ are quantum numbers distinguishing the various resonant states
and $u^r_{st}$ are orthonormal vectors.

Using for states $|in\rangle $ and $|out\rangle $ a partial wave basis
$|j,j^z,\{n\}\rangle $ where $j$ is the total angular momentum, $j^z$ is its
third component and $\{n\}$ is a collection of the other quantum
number characterizing the state, due to angular momentum conservation
we can write
 $$
 u_{j_i, \{n\}}^{j^z_i}=\delta_{j_i, j_{in}}\delta_{j^z_i,
 j^z_{in}}{\it u}_{\{n\}}\ .
 $$

The cross section for a process $a\to b$ is then (see \cite{weinberg},
 Chapter 3)
 \beq 
 \sigma={16\ov s}{\pi \ov \mathcal{N}}\sum_j (2j+1)
  {M_r^2\Gamma_{ja}\Gamma_{jb} \ov (s-M_r^2)^2+\Gamma_j^2 M_r^2}\ ,
 \eeq
where $\mathcal{N}$ is a numerical factor giving the multiplicity of
the initial states and
$\Gamma_{ja(b)}=\Gamma_j\sum_n|{\it u}^{a(b)}_{\{n\}}|^2$
is the partial width corresponding to the decays of resonance $|j\rangle$ into
 states $|a(b)\rangle $.



\begin{thebibliography}{99}  

\bibitem{ADD} 

N.~Arkani-Hamed, S.~Dimopoulos and G.~R.~Dvali,
Phys.\ Rev.\ D {\bf 59}, 086004 (1999)
[arXiv:hep-ph/9807344].

\bibitem{hoyle}
C.D. Hoyle, D.J. Kapner, B.R. Heckel, E.G. Adelberger, J.H. Gundlach,
U. Schmidt, H.E. Swanson (Washington U., Seattle),. 
Phys.Rev.D {\bf 70}, 042004 (2004)
[arXiv:hep-ph/0405262];
C.D. Hoyle, U. Schmidt, B.R. Heckel, E.G. Adelberger, J.H. Gundlach,
D.J. Kapner, H.E. Swanson (Washington U., Seattle),
Phys.Rev.Lett.{\bf 86},1418, (2001)
[arXiv: hep-ph/0011014]

\bibitem{antoniadis}
  I.~Antoniadis, N.~Arkani-Hamed, S.~Dimopoulos and G.~R.~Dvali,
  Phys.\ Lett.\ B {\bf 436} (1998) 257
  [arXiv:hep-ph/9804398].




\bibitem{giudice}
  G.~F.~Giudice, R.~Rattazzi and J.~D.~Wells,
  Nucl.\ Phys.\ B {\bf 544}, 3 (1999)
  [arXiv:hep-ph/9811291].


\bibitem{peskin0} E.~A.~Mirabelli, M.~Perelstein and M.~E.~Peskin,
``Collider signatures of new large space dimensions,''
Phys.\ Rev.\ Lett.\  {\bf 82}, 2236 (1999)
[arXiv:hep-ph/9811337].

\bibitem{dudas}
E.~Dudas and J.~Mourad,
``String theory predictions for future accelerators,''
Nucl.\ Phys.\ B {\bf 575}, 3 (2000)
[arXiv:hep-th/9911019].


\bibitem{peskin} S.~Cullen, M.~Perelstein and M.~E.~Peskin,
``TeV strings and collider probes of large extra dimensions,''
Phys.\ Rev.\ D {\bf 62}, 055012 (2000)
[arXiv:hep-ph/0001166].


\bibitem{hewett} J.~L.~Hewett,
``Indirect collider signals for extra dimensions,''
Phys.\ Rev.\ Lett.\  {\bf 82}, 4765 (1999)
[arXiv:hep-ph/9811356].

\bibitem{bando} M.~Bando, T.~Kugo, T.~Noguchi and K.~Yoshioka,
Phys.\ Rev.\ Lett.\  {\bf 83}, 3601 (1999)
[arXiv:hep-ph/9906549].


\bibitem{AAB}
E.~Accomando, I.~Antoniadis and K.~Benakli,
``Looking for TeV-scale strings and extra-dimensions,''
Nucl.\ Phys.\ B {\bf 579}, 3 (2000)
[arXiv:hep-ph/9912287].



\bibitem{wells} H.~Murayama and J.~D.~Wells,
``Graviton emission from a soft brane,''
Phys.\ Rev.\ D {\bf 65}, 056011 (2002)
[arXiv:hep-ph/0109004].


\bibitem{GS}
  M.~B.~Green and J.~H.~Schwarz,
  Nucl.\ Phys.\ B {\bf 198} (1982) 441.



\bibitem{schwarz}
  J.~H.~Schwarz,
  Phys.\ Rept.\  {\bf 89}, 223 (1982).



\bibitem{HK} A.~Hashimoto and I.~R.~Klebanov,
``Decay of Excited D-branes,''
Phys.\ Lett.\ B {\bf 381}, 437 (1996)
[arXiv:hep-th/9604065];
A.~Hashimoto and I.~R.~Klebanov,
``Scattering of strings from D-branes,''
Nucl.\ Phys.\ Proc.\ Suppl.\  {\bf 55B}, 118 (1997)
[arXiv:hep-th/9611214].


\bibitem{CIR}
 D.~Chialva, R.~Iengo and J.~G.~Russo,
  ``Search for the most stable massive state in superstring theory,''
  JHEP {\bf 0501}, 001 (2005)
  [arXiv:hep-th/0410152].


\bibitem{weinberg}
Steven Weinberg, ``The Quantum Theory of Fields, Vol. 1: Foundations'', 
Cambridge University Press

\bibitem{LEP2} 
A.~Litke, M.~Maggi, G.~Taylor and I.~Tomalin  [ALEPH Collaboration],
``Search for extra spatial dimensions and TeV scale quantum gravity at  LEP,''
CERN-OPEN-99-269
{\it Prepared for International Europhysics Conference on High-Energy Physics (EPS-HEP 99), Tampere, Finland, 15-21 Jul 1999}.



\end{thebibliography}
\end{document}